# A Practical Method for Pupil segmentation in challenging conditions

[1]Donya Khaledyan *, [1] Mohammad Eshghi, [2]Morteza Heidari,[3]Abolfazl Zargari Khuzani, [4]Najmeh Mashhadi
[1]Department of Electrical Engineering, Shahid Beheshti University, Tehran, Iran.
[2]School of Electrical & Computer Engineering, University of Oklahoma, Norman, USA
[3]The Department of Electrical and Computer Engineering, University of California, Santa Cruz, USA
[4]The Department of Computer Science and Engineering, University of California, Santa Cruz, USA
* d.khaledyan@mail.sbu.ac.ir

*Abstract*— **Various methods have been proposed for authentication, including password or pattern drawing, which is clearly visible on personal electronic devices. However, these methods of authentication are more vulnerable, as passwords and cards can be forgotten, lost, or stolen. Therefore, a great curiosity has developed in individual authentication using biometric methods that are based on physical and behavioral features not possible to forget or be stolen. Authentication methods are used widely in portable devices since the lifetime of battery and time response are essential concerns in these devices. Due to the fact that these systems need to be fast and low power, designing efficient methods is still critical. We, in this paper, proposed a new low power and fast method for pupil segmentation based on approximate computing that under trading a minor level of accuracy, significant improvement in power assumption and time saving can be obtained and makes this algorithm suitable for hardware implementation. Furthermore, the experimental results of PSNR and SSIM show that the error rate in this method is negligible.**

## I. INTRODUCTION

The biometric characteristics are unique and not possible to change. Each one of the biometric methods has a different level of security. Iris is one of the ways with a high level of protection [1]. In the present era, power consumption is a significant concern, especially in electronic systems, therefore applying the power reduction techniques and implementing low power circuits and systems attract much attention in recent years [2-6]. The power consumption can be reduced by applying different low power techniques [7]. These techniques include emerging technologies [8], gate, and RTL methods such as clock and data gating [7, 9], algorithm level methods like approximate computing [10, 11], and system levels techniques such as dynamic voltage scaling and frequency governing [7].

Approximate computing on images has recently appeared as a capable approach to design energy-efficient systems. Han and Orshansky in [12] investigate the recent developments in the area of approximate computing. Especially in error resilient applications like multimedia (signal, image, and video) processing. Notably, the final output of the iris segmentation system is an image; this provides the output of this system to be numerically approximate.

The progress of low cost, small size, and low power embedded systems for iris and pupil recognition is a hot topic of research in recent years [13]. In [14], an edge-map extraction technique for pupil detection in the Near Infra-Red (NIR) images is presented. In these types of images, the illumination is controlled. Near-infrared illumination also helps to expose the detailed construction of severely pigmented (dark) irises.

In general, iris and pupil segmentation methods are categorized into two main classes: Circular Hough Transform (CHT) methods [15], and gradient-based methods [16]. There are also other techniques, such as methods that are mixtures of both circular and gradient methods and other methods such as learning-based methods and histogram-based methods.

The most significant work in the history of iris biometrics is Daugman's patent [16]. It is reasonable to say that iris biometrics has advanced with the concepts in Daugman's method. This work is becoming a standard reference model. Daugman's approach uses a gradient-based method. This paper expresses gradient-based methods are more straightforward, faster, and more suitable for real-time and fast applications. We will use this idea to present the gradient-based edge detection algorithm.

Wildes's proposed method [17] is also one of the notable works in the iris and pupil recognition field. Wildes has used circular transform.

A new idea is presented in [18] based on a combination of gradient and circular edge detection methods to improve the efficiency of pupil segmentation in comparison with the Daugman's and Wildes's techniques. Their results are compared with the prior methods; both accuracy and execution time are enhanced. Furthermore, in multimedia processing (audio, video, and image), pattern recognition and data mining sometimes accurate results are not necessary and approximate results can work fine. For example, sharpness, resolution, and colours are not critical to them. These applications are called

approximate resilient. Sources of approximate resilient can be divided into two categories:

1) Perceptual limitations due to the ability of the human brain to fill the missing information, and human eye limitation to obtain and distinguish high-frequency patterns.

2) Redundant input: an algorithm can be lossy and still provide acceptable results.

For example, as stated in [19] and shown in Fig.1, by accepting a 6.25% error, the overall performance of their proposed system is improved 16 times. In these systems, with a combination of efficient hardware, and software, approximate computing are promising approaches for designing energy-efficient computing methods. For instance, an approximate multi-bit adder can be implemented using a single bit accurate full-adders for more significant bits and approximate full-adders for less significant bits.

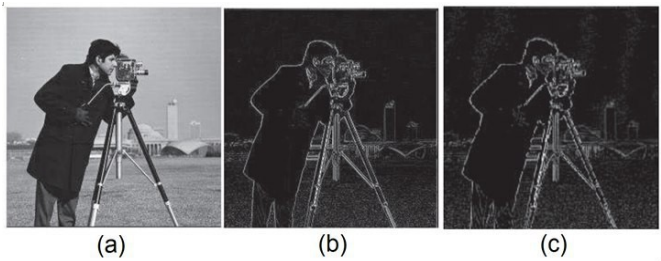

Fig. 1. An example of approximate computing [19] a) original grayscale image b) accurate edge detection c) approximate edge detection using truncation with considering first 4 MSBs.

From the hardware point of view, approximate circuits usually use fewer transistors and occupy less area on chips. Furthermore, the approximate circuits are usually faster than the accurate circuits because in the approximate circuits, the paths between inputs and outputs are shorter, and capacitance load is generally smaller. Hence, when it is not necessary to achieve a high level of accuracy, and power, speed are more critical, the important of the approximate computing felt better.

The rest of the paper is organized as follows: in Section II some background of pupil segmentation, including Gaussian and Sobel filter, image binarization and approximate computing alongside the related works are discussed. The proposed method for power and hardware reduction is presented in section III. Simulation results of the proposed method presented in section IV, finally based on the simulation results section V concludes the paper.

## II. BACKGROUND

Iris recognition consists of 2 main steps [20]. First, the pupil and iris will be detected in the eye image. Then in the second step, features of iris will be extracted. Accuracy in assigning the actual pupil and iris boundaries, even if they are partially invisible, is essential because the mapping of the iris in an iris recognition system is critically dependent on this step. Inaccuracy in the segmentation step can cause different mappings of the pupil and iris pattern, and such deviations could cause failures to feature extraction. Otherwise, the first step needs more time for localizing the pupil, because the whole image of the eye must be processed to determine the pupil.

As we described in [14] an edge-map extraction technique for pupil detection in the NIR images is presented. Their method significantly improved the accuracy of the result by reducing the extracted false edges. One of the advantages of this method is that it is applicable to both constrained and limited constrained images (images that are affected by noises such as eyelashes, eyelids, the reflection of light, and eyebrow [21]) NIR images. This method is a combination of two other methods (thresholding and edge detection) to evaluate the pupil image. In our approach, we likewise used the combination of thresholding and edge detection method to take the advantages of the method presented in [14]. Fig.2 shows the method presented in this paper. All parts of this figure described in the preceding sections. First, by applying the Gaussian filter, high-frequency noise is eliminated from the image, then by applying each one of the two methods mentioned above (thresholding and edge detection), pupil will be identified precisely. Here we will describe details of both Gaussian and Prewitt filters as two essential steps in this process.

### A. Gaussian filter

The Gaussian filter has been widely used in image processing and computer vision. Gaussian filter is practical for noise detention and image smoothing as pre-processing steps [22]. The two-dimensional digital Gaussian filter is expressed as:

$$G(x,y) = \frac{1}{2\pi\sigma^2} e^{-\frac{(x^2+y^2)}{2\sigma^2}} \quad (1)$$

$$K = \frac{1}{16}\begin{bmatrix} 1 & 2 & 1 \\ 2 & 4 & 2 \\ 1 & 2 & 1 \end{bmatrix} \quad (2)$$

The image is characterized as a 2D array where $x$ and $y$ are the row and column indices, and $\sigma^2$ is the variance of Gaussian filter. For simplicity and optimality of the hardware implementation, standard deviation ($\sigma$) and the dimension of the Gaussian filter are considered 1 and 3*3 respectively. The kernel of the Gaussian filter is expressed by Eq. (2). The Eq. (3) is the result of convolving input image ($I$) with $K$.

$$I_O = K*I \quad (3)$$

The block diagram of the Gaussian filter is shown in Fig. 3. In this architecture, as shown, shifting is used to implement multiplication and division operations. This idea will preserve time, which indicates for hardware implementation purposes, this algorithm needs less time and resources in comparison with implementing the algorithm by multiplication.

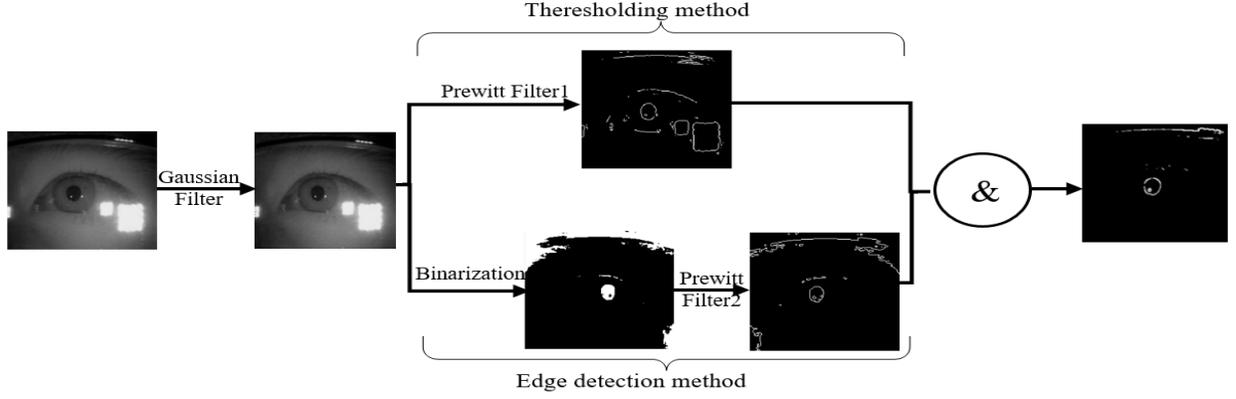

Fig. 2. The block diagram of proposed method

## B. Prewitt filter

The Prewitt filter is an edge detection filter. The edge detection filters are generally based on the gradient. In an image, the edge is where discontinuity in the intensity of the pixels occurs. This filter has two kernels $k_x$ and $k_y$ to find x and y derivative individually. These kernels are represented by Eq. (4) and (5). $G_x$ and $G_y$, which are the results of convolving kernels with image, will be computed using Eqs. (6) and (7). Then the two separate images result from $G_x$ and $G_y$ are combined by using Eq. (8) the local edges are obtained by this equation.

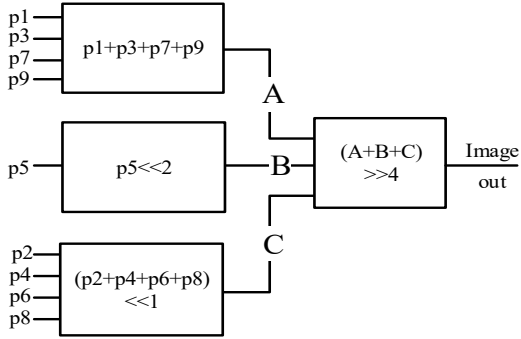

Fig. 3. Gaussian filter block diagram [14]

However, it will be expensive and time-consuming to calculate square and square root operations for every pixel, therefore an approximate equation $G=[G_x]+[G_y]$ as shown in Fig.4 is used to reduce the processing time and hardware resources.

$$k_x = \begin{bmatrix} -1 & 0 & 1 \\ -1 & 0 & 1 \\ -1 & 0 & 1 \end{bmatrix} \quad (4)$$

$$k_y = \begin{bmatrix} -1 & -1 & -1 \\ 0 & 0 & 0 \\ 1 & 1 & 1 \end{bmatrix} \quad (5)$$

$G_x = p3+p6+p9-p1-p4-p7$ (6)

$G_x = p7+p8+p9-p1-p2-p3$ (7)

$$GM = \sqrt{Gx^2 + Gy^2} \quad (8)$$

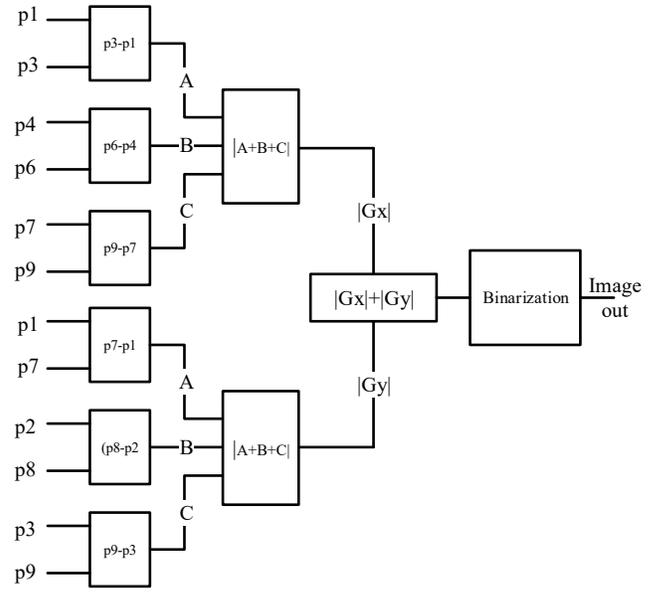

Fig. 4. Prewitt filter block diagram

## C. Image binarization:

Image binarization takes a grayscale image (0 to 255 grayscale levels) and converts it to black-and-white with pixel values of (0 or 1). This is a task frequently executed when trying to extract an object from an image. The binary value of each pixel is $b(x, y)$, and the intensity of pixels in the main image is $f(x, y)$. A threshold value is required to binarize the image. In [19, 23] to present an acceptable threshold, they first obtain the minimum pixel value of the image, and by adding with 40, the threshold is calculated, since in the iris images, the minimum value of the pupil is 0; therefore, the threshold value will be 40, but we analysed the images and evaluated the results. It showed threshold=40 does not work well, then in a heuristic way concluded that for the thresholding step, the best value for the image binarization would be 96. Image binarization is performed as:

$$b(x,y) = \begin{cases} 1 & f(x,y) > T \\ 0 & O.W \end{cases} \qquad (9)$$

## III. PROPOSED METHODS TO REDUCE POWER, HARDWARE AND TIME

As mentioned before, approximate computing can cause a significant drop in hardware complexity, power consumption, and time. In this section, some serviceable techniques to the proposed architecture are presented, which lead to a significant reduction in time processing, besides these techniques make this process more efficient for hardware implementation.

### A. Using approximate adder with accurate final results:

In the Gaussian filter, the final value of convolving image with Gaussian kernel, as shown in Eq. (2) is divided by 16 (done by four shifts to the right). Therefore, four least significant bits are removed. Hence, approximate units can be used for the calculation of these bits. Due to this fact, the approximated units just need to compute carry out. Already stated that these bits are eliminating, so as a result, this method does not cause any error.

### B. Using approximate adder with approximate final results:

By using four approximate full-adder in the previous subsection, the result is still accurate. With accepting small inaccuracy and employing more than four approximate full-adder, more power and area-saving can be obtained. It should be noted that applying these techniques cause the least error in the output compared to the result without applying techniques.

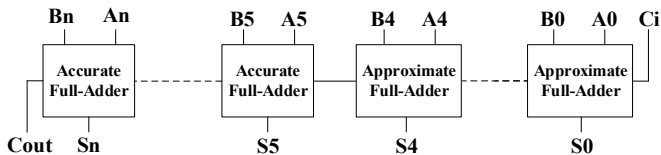

Fig. 5. Employed multi bit approximate adder in Gaussian and Prewitt filters

### C. Simplified image binarization:

In the binarization step, by setting a sensible threshold value, some LSBs can be ignored, simulation results show that threshold value is equal to 96 (01100000). In the image binarization step, 5 Least Significant Bits (LSBs) can be neglected while an acceptable output guaranteed.

In Prewitt filter, the threshold value was assumed to be 128 (000100000000); therefore, the 8 bit can be ignored.

By applying this technique, the volume of the comparator circuit hardware is diminished more than 50%; as a result, power consumption is significantly reduced, due to the shortening of the critical path, the delay is reduced as well.

### D. Multi bit approximate full-adder:

The Gaussian and Prewitt filters are supposed to be based on the proposed approximate multi-bit adder circuit shown in Fig.5. This circuit employed approximate FAs for 5 LSBs and accurate FAs for 3 MSBs for Gaussian and 7 MSBs for Prewitt.

### E. Simplified Prewitt filter2:

According to Fig. 2, before the Prewitt filter2 is the image binarization, therefore, the input image of this part is binarized and the binarization part of Prewitt filter2 is not necessary; therefore, it can be deleted. This will reduce hardware, power, and time consumption of the system.

## IV. RESULTS

To verify the functionality of the proposed method, some MATLAB simulations carry out using CASIA Iris-Thousand, version 4.0 (CITHV4); and CASIA Iris-Lamp, version 3.0 (CILV3) database which contains NIR image databases consist of less constrained and constrained images are utilized. These databases are freely available on the internet at [24].

Fig. 6 shows the simulation results of the proposed method in comparison with the accurate method in [14]. As it stands, the proposed method has very similar results in comparison with the accurate method. However, as discussed, with the techniques proposed in our paper, power and area will be saved, which performs this technique suitable for hardware implementation. In Fig. 6 the unconstrained images are shown to prove that the technique used in this paper for pupil segmentation is suitable for these types of images. The mean value for image quality metrics, the peak signal-to-noise ratio (PSNR), and structural similarity index (SSIM) are measured on the database for images that exact method is applied and images that proposed techniques are used, the mean of PSNR for these datasets is 26.90125. The mean of SSIM is 0.989025. The results firmly indicate that the proposed method has very similar results in comparison with accurate methods. At the same time, circuit complexity, power consumption, and area are reduced significantly.

To calculate the execution time of each method, the MATLAB's timer functions 'tic' and 'toc' is utilized. Table 1 demonstrate that the proposed method results enhanced pupil segmentation accuracy and execution time due to reducing wrong edges in the edge-maps of eye images.

## V. CONCLUSION AND FUTURE SCOPE

Low cost, low power, and fast response time are critical parameters for practical authentication systems. The primary purpose of the work described in this study was to providing techniques to reduce power, area, and time in a specific pupil segmentation system. We presented a method that reduces the wrong edges. Then, by taking advantage of approximate computing and a simplified comparator, a new architecture for pupil segmentation is proposed, which is suitable for low power and fast applications. Simulation results show that by accepting a minor level of inaccuracy in comparison with accurate methods, considerable area and power saving can be obtained. In the near future, based on the method and results of this paper, the hardware implementation on the FPGA platform will be developed, in order to perform a real-time pupil segmentation, also on FPGA we can report the amount of power that will decrease based on the proposed method.

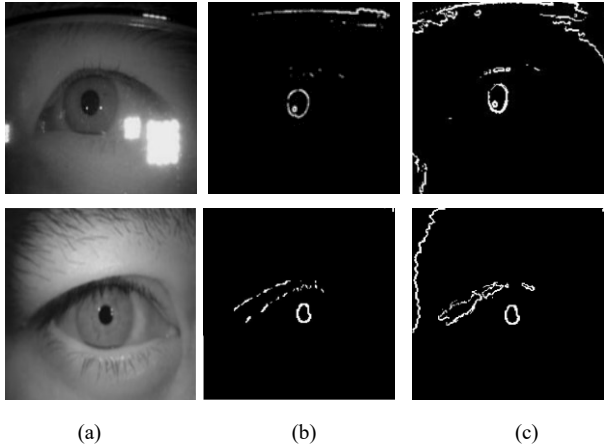

(a)          (b)          (c)

Fig. 6. Functional validation of the proposed method in comparison to [14]. a) image test b) accurate results c) approximate results of proposed method.

Table1. Time and accuracy comparison with similar algorithms

| | Method | Execution time | Accuracy (%) |
|---|---|---|---|
| Zaim [18] | Gradient | 5.83 s | 92.7 |
| Kennedy [15] | Circular | 9.02 s | 93 |
| Kumar [14] | Gradient | 1.90 s | 94 |
| **Proposed** | Gradient | 1.12 s | 92.5 |